\documentstyle[prb,aps,epsfig,floats,goodfloat]{revtex}

\newcommand{\bc}{\begin{center}}
\newcommand{\ec}{\end{center}}
\newcommand{\be}{\begin{equation}}
\newcommand{\ee}{\end{equation}}
\newcommand{\beqn}{\begin{eqnarray}}
\newcommand{\eeqn}{\end{eqnarray}}
\newcommand{\ql}{q_{l}}
\newcommand{\ml}{\mu_{l}}
\newcommand{\av}{_{\mathrm{av}}}
\begin{document}
\draft

\twocolumn[\hsize\textwidth\columnwidth\hsize\csname@twocolumnfalse%
\endcsname

\title{
Nature of the Spin Glass State
}

\author{Matteo Palassini and A. P. Young}
\address{Department of Physics, University of California, Santa Cruz, 
CA 95064}

\date{\today}

\maketitle

\begin{abstract}
The nature of the spin glass state is investigated by studying changes to the
ground state when a weak perturbation is applied to the bulk of the system. We
consider short range models in three and four dimensions and the
infinite range Sherrington-Kirkpatrick (SK) and Viana-Bray models. 
Our results for the SK and Viana-Bray models agree with the replica 
symmetry breaking picture.
The data for the short range models fit naturally a picture in which
there are large scale excitations which cost a finite energy but whose
surface has a fractal dimension, $d_s$, less than the space dimension $d$.  
We also discuss a possible crossover to other behavior at
larger length scales than the sizes studied. 
\end{abstract}

\pacs{PACS numbers: 75.50.Lk, 05.70.Jk, 75.40.Mg, 77.80.Bh}
]

The nature of ordering in spin glasses below the transition temperature, $T_c$,
remains a controversial issue.  Two theories have been extensively discussed: the
``droplet theory'' proposed by Fisher and Huse\cite{fh} (see also
Refs.~\onlinecite{bm,mcmillan,ns}), and the replica symmetry
breaking (RSB) theory of Parisi\cite{parisi,mpv,by}. An important difference
between these theories concerns the number of large-scale, low energy
excitations. In the RSB theory, which follows the exact solution of the
infinite range SK model, there are excitations which involve turning over a
finite fraction of the spins and which cost only a {\em finite} 
energy even in the thermodynamic limit.  Furthermore, the surface of these
excitations is argued\cite{qlink} to be
space filling, i.e. the fractal
dimension of their surface, $d_s$, is equal to the space dimension, $d$.
By contrast, in the droplet theory, the lowest energy excitation which involves
a given spin and which has linear spatial extent $L$ typically costs an energy of order
$L^\theta$, where $\theta$ is a (positive) exponent. Hence, in the
thermodynamic limit, excitations which flip a finite fraction of the spins cost
an {\em infinite} energy. Also, the
 surface of these excitations is not
space filling, {\em i.e.} $d_s < d$.

Recently we\cite{py1,py2,py3} investigated this issue by looking at how spin
glass ground states in two and three dimensions change upon changing the
boundary conditions. Extrapolating from the range of sizes studied to the
thermodynamic limit, our results suggest that the low energy
excitations have $d_s < d$. Similar results were found in two dimensions
by Middleton\cite{midd}. In this paper, following a suggestion by
Fisher\cite{dsf}, we apply a perturbation to the ground states in the {\em
bulk} rather than at the surface. The motivation for this is two-fold:
(i) We can apply the same method both to models with short range interactions and
to infinite range models, like the SK model,
and so can verify that the method is able to distinguish
between the RSB picture, which is believed to apply to infinite range models,
and some other picture which may apply to short range
models.
(ii) It is possible that there are
other low energy excitations which are not excited by changing the
boundary conditions\cite{kawa,hm}. 

We consider the short-range Ising spin glass in three and four
dimensions, and, in addition, the SK and Viana-Bray\cite{vb}
models. The latter is
infinite range but with a finite average
coordination number $z$, and is expected to show RSB behavior.
All these models have a finite
transition temperature. 

Our results for the SK and Viana-Bray models
show clearly the validity of the RSB picture. 
However, for
the short range models, our data is consistent with a picture suggested by
Krzakala and Martin\cite{km} where there 
are extensive excitations with {\em finite} energy, i.e. their energy varies
as $L^{\theta'}$ with $\theta' = 0$,
but $d_s < d$. In three dimensions, this picture is difficult to
differentiate from the droplet picture where the energy varies as
$L^\theta$, because of the small value of $\theta$ ($\simeq 0.2$,
obtained from
the magnitude of the change of the ground state energy when
the boundary conditions are changed from periodic to anti-periodic\cite{theta-3d}). 
It is easier to
distinguish the two pictures in 4-D, even though the range of $L$ is less,
because $\theta$ is much larger\cite{theta-4d} ($\simeq 0.7$).


The Hamiltonian is given by
\begin{equation}
{\cal H} = -\sum_{\langle i,j \rangle} J_{ij} S_i S_j ,
\label{ham}
\end{equation}
where, for the short range case, the sites $i$ lie on a 
simple cubic lattice in dimension $d=3$ or 4 with $N=L^d$ sites 
($L \le 8$ in 3-D, $L \le 5$ in 4-D), $S_i=\pm
1$, and the $J_{ij}$ are nearest-neighbor interactions chosen from a
Gaussian distribution with zero mean and standard deviation unity. Periodic
boundary conditions are applied. For the SK
model there are interactions between {\em all} pairs chosen from a Gaussian
distribution of width $1/\sqrt{N-1}$, where $N \le 199$. For the Viana-Bray
model each spin is connected with $z=6$ spins on average, chosen randomly,
the width of the Gaussian distribution is unity, and
the range of sizes is $N \le 399$.
To determine the ground state
we use a hybrid genetic algorithm introduced by Pal\cite{pal}, as
discussed elsewhere\cite{py2}.

Let $S_i^{(0)}$
be the spin configuration in the ground state for
a given set of bonds.
Having found $S_i^{(0)}$, we
then add a perturbation to the Hamiltonian designed to
increase the energy of the ground state relative to the other states, and so
possibly induce a change in the ground state. This perturbation,
which depends upon a positive parameter $\epsilon$, changes the
interactions $J_{ij}$ by an amount proportional to $S_i^{(0)} S_j^{(0)}$, i.e.
\begin{equation}
\Delta {\cal H}(\epsilon)  =  \epsilon {1 \over N_b} \sum_{\langle i,j \rangle}
S_i^{(0)} S_j^{(0)} S_i S_j,
\end{equation}
where 
$N_b$ is the number of bonds in the Hamiltonian.
The energy of the ground state will thus increase exactly by an amount
$ \Delta E^{(0)} = \epsilon .$
The energy of any other state, $\alpha$ say, will increase by the lesser amount
$ \Delta E^{(\alpha)} = \epsilon\ \ql^{(0, \alpha)},$
where $\ql^{(0, \alpha)}$ is the ``link overlap'' between the states
``0'' and $\alpha$, defined by
\begin{equation}
\ql^{(0, \alpha)} = {1 \over N_b}\sum_{\langle i,j \rangle} S_i^{(0)} S_j^{(0)} 
S_i^{(\alpha)} S_j^{(\alpha)} ,
\end{equation}
in which the sum is over all the $N_b$ pairs where there are interactions.
Note that the {\em total} energy of the states is changed by an
amount of order unity.

The decrease in the energy {\em difference} between 
a low energy excited state and the
 ground state is given by
\begin{equation}
\delta E^{(\alpha)} = \Delta E^{(0)} - \Delta E^{(\alpha)} = 
\epsilon \ (1 - \ql^{(0, \alpha)}) .
\label{de}
\end{equation}
If this exceeds the original difference in energy, $E^{(\alpha)} - E^{(0)}$,
for at least one of the excited states, then the ground state will change 
due to the perturbation. We denote the new ground state spin
configuration by $ \tilde{S}_i^{(0)}$, and indicate by
$\ql$ and $q$, with no indices, the link- and spin-overlap
 between the new and old ground states.

\begin{figure}
\begin{center}
\epsfig{figure=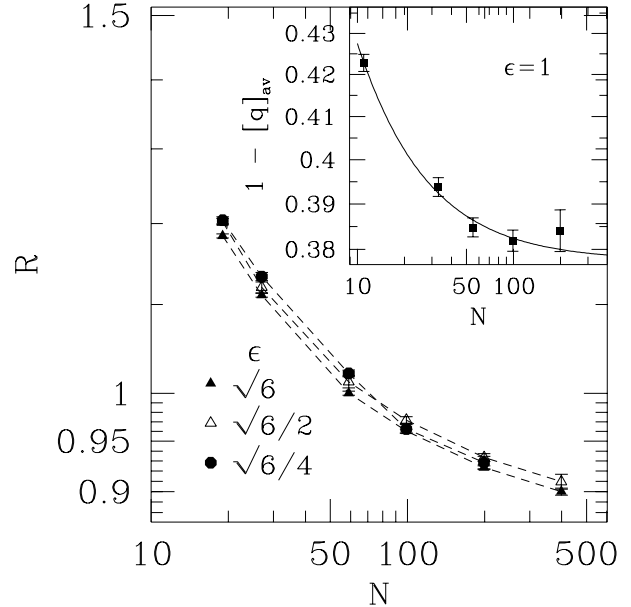,width=\columnwidth}
\end{center}
\caption{
Data for
$R = (1-[q_l]\av)/(1-[q]\av)$ for
the Viana-Bray model with coordination number $z=6$ for several values of
$\epsilon$. The curvature is a strong indication that the data tends to a
non-zero value for $N \to \infty$, as for the SK model. The best fits
to $a + b/N^c$, shown by the lines, give
$a= 0.872 \pm 0.005, 0.883 \pm 0.01$ and $0.84  \pm 0.03$
for $\epsilon=\sqrt{6}, \epsilon=\sqrt{6}/2$ and $\epsilon=\sqrt{6}/4$
respectively.
Inset: 
$1 - [q]\av$ for the SK model
with the strength of the
perturbation given by $\epsilon = 1$.
Because $\ql = q^2$, the behavior of $[\ql]\av$ is
very similar. The data is clearly tending to a constant at large $N$. The solid
line is the best fit 
and has
$a = 0.377 \pm 0.004$.
}
\label{q_sk}
\end{figure}

Next we discuss the expected behavior of $q$ and $\ql$ for
the various models. For the SK
model, it is easy to derive the trivial relation,
$\ql = q^2$ (for large $N$).
Since RSB theory is expected to be correct, 
there are some excited states
which cost a finite energy and which have an overlap $q$ less than
unity. According to Eq.~(\ref{de}), these have a finite
probability of becoming the new ground state. Hence the average value of 
$q$ and $\ql$ over many samples,
denoted by $[\cdots]\av$,
should tend to a constant less than unity in
the thermodynamic limit. 
This behavior is shown in the inset of Fig.~\ref{q_sk}. For the
Viana-Bray model, where there is no trivial connection between $q$ and $\ql$,
we show in Fig.~\ref{q_sk}
data for $R = (1-[\ql]\av)/(1-[q]\av)$ for several values of $\epsilon$.
This also appears to saturate. 
We plot this ratio rather than $[q]\av$
or $[q_l]\av$ for better comparison with the short range case below.
For both models we took
$\epsilon$ to be a multiple of
the transition temperature (the mean field
approximation to it, $T_c^{MF} = \sqrt{z}$,
for the Viana-Bray model), so that a perturbation of
comparable magnitude was applied in both cases.

\begin{figure}
\begin{center}
\epsfig{figure=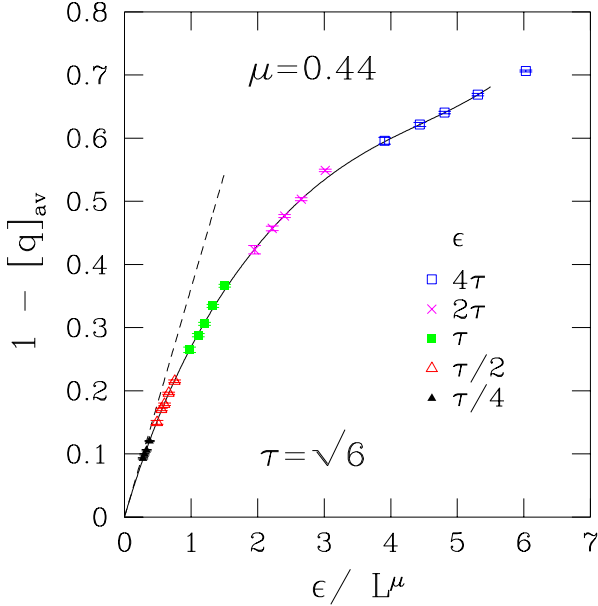,width=\columnwidth}
\end{center}
\caption{
A scaling plot of the data for $[q]\av$ in 3-D according to
Eq.~(\ref{scaling}). The data collapse is very good with $\mu = 0.44$.
The solid curve is a polynomial fit ($\chi^2=14.9$, $d.o.f.=13$), constrained to go through the origin,
omitting the $L=3$ data. The dashed line is the linear term in the fit.
}
\label{q_scaling_3d}
\end{figure}

What do we expect for the short range models? In the RSB 
theory,  $1-[q]_{av}$ and $1-[q_l]_{av}$ (and hence the
ratio $R$) should saturate to a finite value for large $L$.
To derive the prediction of the droplet theory,
suppose that the energy to create an excitation of linear dimension, $l$,
has a characteristic scale of $l^{\theta'}$
(we use $\theta'$ rather than
$\theta$ to allow for the possibility that this exponent is different from the
one found by changing the boundary conditions). Let us assume that large clusters
($l \approx L$) dominate and ask for
the probability that a large cluster is excited.
The energy gained from the perturbation is 
$\epsilon (1 - \ql) \sim \epsilon / L^{(d - d_s)}$ since $1/L^{(d-d_s)}$ is the
fraction of the system containing the surface (i.e. the broken bonds) of the
cluster. Generally this will not be able to overcome the $L^{\theta'}$ energy
cost to create the cluster. However, there is a distribution of
cluster energies and if we make the plausible hypothesis that this
distribution  has a finite weight at the origin, then the
probability that the cluster is excited 
is proportional to
$1/L^{d - d_s + \theta'}$. In other words
\begin{equation}
1 - [q]\av \sim \epsilon / L^{\mu} \ \ {\mathrm where} \ \ 
\mu = \theta' + d - d_s .
\label{psis}
\end{equation}
As discussed above, $1 - \ql$ is of order $1/L^{(d-d_s)}$ 
and so
\begin{equation}
1 - [\ql]\av \sim \epsilon / L^{\ml} \ \ {\mathrm where} \ \ 
\ml = \theta' + 2(d - d_s) .
\label{psil}
\end{equation}
Similar expressions have been derived by Drossel et al.\cite{drossel} in
another context.
Eqs.~(\ref{psis}) and
(\ref{psil}) are expected to be valid only {\em asymptotically} 
in the limit $\epsilon \to 0$. In order
to include data for a range of values of $\epsilon$ we note that the data is
expected to scale as 
\begin{eqnarray}
1 - [q]\av & = & F_q(\epsilon/L^\mu) ,  \nonumber \\
 1 -  [\ql]\av & = & L^{-(d-d_s)}
F_{q_{l}}(\epsilon/L^\mu) ,
\label{scaling}
\end{eqnarray}
where the scaling functions $F_q(x)$ and $F_{q_{l}}(x)$
both vary linearly for small $x$. Note that the above discussion
applies also to a picture in which $\theta'=0$ and $d_s<d$.


\begin{figure}
\begin{center}
\epsfig{figure=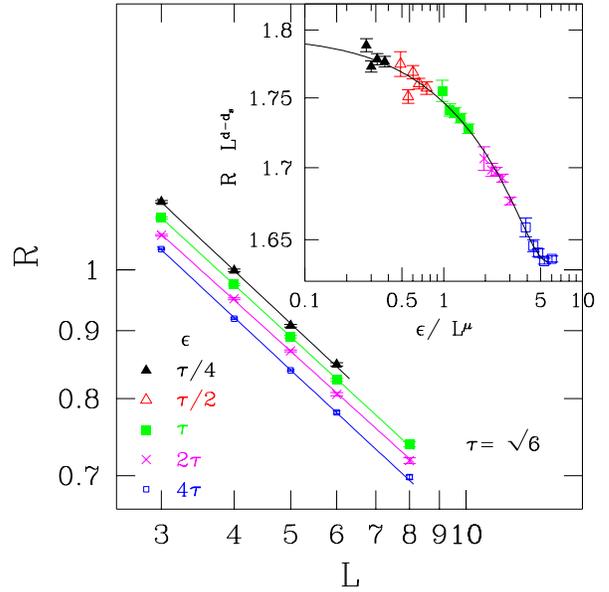,width=\columnwidth}
\end{center}
\caption{
A plot of $R = (1-[q_l]\av)/(1 - [q]\av)$, the surface to volume ratio of the
clusters, in 3D as a function of system size for different different values of 
$\epsilon$. For clarity the $\epsilon = \tau/2$ data is omitted.
The dependence on $\epsilon$ is quite weak, and for
each value of $\epsilon$ the data gives a good fit to a straight line with
slope [equal to $-(d-d_s)$] consistent with Eq.~(\ref{ds}).
The inset shows a scaling plot
of the data according to Eq.~(\ref{scaling}) with the same value of $\mu$ as
in Fig.~\ref{q_scaling_3d}. The solid curve is a polynomial fit 
($\chi^2=26.0$, $d.o.f.=18$). 
}
\label{q_ql_3d}
\end{figure}

A scaling plot of our results for $1 - [q]\av$ in 3D
is shown in
Fig.~\ref{q_scaling_3d}.
We consider a range of $\epsilon$ from $\sqrt{6}/4$  to
$4\sqrt{6}$ (note that $T_c^{MF}= \sqrt{6}$) and find that the data collapse
well onto the form expected in Eq.~(\ref{scaling}) with
$ \mu = 0.44 \pm 0.02.$

It is also convenient to plot the ratio $R$, which
represents the surface to volume ratio of the excited clusters. This has
a rather
weak dependence on $\epsilon$ and, as shown in Fig.~\ref{q_ql_3d},
the data for {\em each}\/ of the values of $\epsilon$
fits well the power law behavior
$L^{-(d-d_s)}$, expected from Eqs.~(\ref{psis}) and (\ref{psil}), with
$d - d_s$ between 0.40 and 0.41 (the goodness of fit parameter, $Q$,
is $0.07, 0.03, 0.85, 0.23, 0.10$, in order of increasing $\epsilon$).
The inset to Fig.~\ref{q_ql_3d} shows that there are 
small deviations from the asymptotic behaviour, which can be accounted for
by a scaling function with the same value of $\mu$ as in
Fig.~\ref{q_scaling_3d} and with
\begin{equation}
d - d_s = 0.42 \pm 0.02 \ \ (3D) .
\label{ds}
\end{equation}
From this value of $\mu$ and
Eqs.~(\ref{psis}) and (\ref{ds})
we find
\begin{equation}
\theta' = 0.02 \pm 0.03 \ \ (3D) .
\end{equation}
In order to test the RSB prediction, we tried fits of the form 
$ R = a + b/L^c $, which give
$a =  0.28 \pm 0.18, 0.01 \pm 0.14, 0.04 \pm 0.11$, and $-0.28 \pm 0.18$
($Q=0.08,0.01,0.72$, and 0.52)
for $\epsilon/\tau = 0.25, 0.5, 1$ and $2$. These are consistent with $a=0$
though a fairly small positive value, which would imply
$d_s=d$, cannot be ruled out. For $\epsilon/\tau
=4$ the fit gives a small positive value, $0.18 \pm 0.07$ ($Q=0.79$),
but this is likely too large a value of $\epsilon$ to be in the 
asymptotic regime for these sizes (see the inset of Fig.~\ref{q_ql_3d}).
The form $R=a+b/L+c/L^2$ also fits reasonably well the data and gives 
$a$ between 0.41 and 0.48 ($Q=0.16,0.03,0.82,0.80,0.16$).
However, for both forms the data are very far from the asymptotic limit $R\sim a$ 
for the sizes considered, unlike for the Viana-Bray model 
(compare the main parts of Figs. ~\ref{q_sk} and
~\ref{q_ql_3d}). 
By contrast, the deviation from the asymptotic behavior 
$R\sim L^{-(d-d_s)}$ is quite small (see the inset of
Fig.~\ref{q_ql_3d}). 
\begin{figure}
\begin{center}
\epsfig{figure=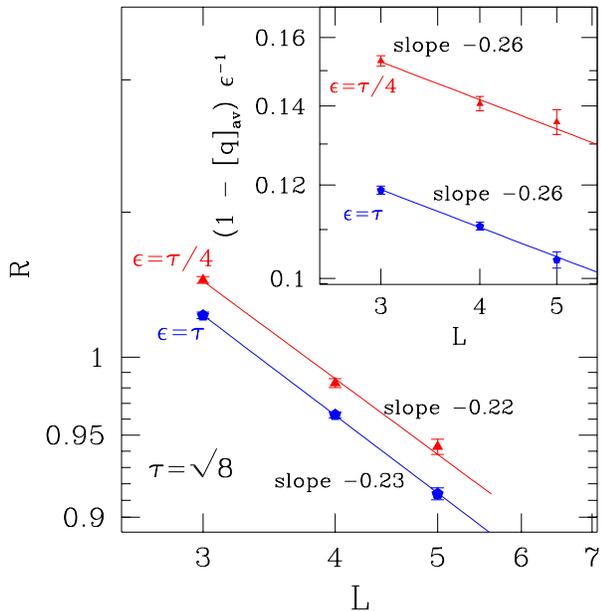,width=\columnwidth}
\end{center}
\caption{
The main figure shows the ratio $R = (1-[q_l]\av)/(1-[q]\av)$
for different values of $\epsilon$
in 4-D. The inset shows corresponding data for $[q]\av$.
}
\label{q_all_4d}
\end{figure}

In Fig.~\ref{q_all_4d} we show analogous results
in 4-D. The calculations were performed for two
different values $\epsilon = \sqrt{8}/4$ and $\sqrt{8}$ ($= T_c^{MF}$).
The exponents are essentially the same for these two values of the
perturbation and the fits give
$ \mu = 0.26 \pm 0.04$, $d - d_s = 0.23 \pm 0.02$ ,
and so from Eq.~(\ref{psis}) we get our main results for 4D:
\begin{equation}
\theta' = 0.03 \pm 0.05, \quad d - d_s = 0.23 \pm 0.02 \ \ (4D).
\end{equation}
The data in Fig.~\ref{q_all_4d} is {\em consistent} with the scaling form in
Eq.~(\ref{scaling}) but the data for the two values of $\epsilon$ are too
widely separated to {\em demonstrate} scaling.

Interestingly, our results in both 3-D and 4-D are consistent with $\theta' =
0$, and, within the error bars, (which are purely statistical) incompatible
with the relation $\theta' = \theta$, since
$\theta \simeq 0.20$ in
3-D\cite{theta-3d,py2} and
$\theta \simeq 0.7$ in 4-D\cite{theta-4d}. In 3-D, 
$\theta - \theta'$ is small, but in 4-D this difference is larger and hence the
conclusion that $\theta' \ne \theta$ is stronger. However, the conclusion that
$d-d_s > 0$ is less strong in 4-D because our value for $d-d_s$ is quite small
and the range of sizes is smaller than in 3-D.

It would be interesting, in future work, to study the
nature of these excitations to see
how they differ from the
excitation of energy $L^\theta$ (with $\theta > 0$) induced by boundary
condition changes\cite{py2,theta-3d,theta-4d}.
In particular, if their
volume is space filling,
one would 
expect a non-trivial order parameter distribution, $P(q)$, at finite
temperatures. 

To conclude, an
interpretation of our results for short range models
which is natural, in that it fits the data with a minimum number of
parameters and with small corrections to scaling, is that there
are large-scale low energy excitations which cost a finite energy, and
whose surface has fractal dimension
less than $d$. This picture differs from the one suggested by 
Houdayer and Martin\cite{hm}, in which $d_s=d$. 
Furthermore, the results for short range models
appear quite different from those of the mean-field
like Viana-Bray model for samples with a similar coordination number and
a similar number of spins.
Other scenarios,
such as the droplet theory (with $\theta^\prime = \theta \ (> 0)$) or
an RSB picture (where $\theta^\prime = 0, d - d_s = 0$), require larger
corrections to scaling, but we cannot rule out the possibility of
crossover to one of these behaviors at larger sizes.

We would like to thank D.~S.~Fisher for suggesting this line of enquiry, and
for many stimulating comments. We also acknowledge useful discussions and
correspondence with G.~Parisi, E. Marinari,
O.~Martin, M.~M\'ezard and J.-P.~Bouchaud. We are grateful to D.~A.~Huse,
M.~A.~Moore and
A.~J.~Bray for suggesting the scaling plot in Fig.~\ref{q_scaling_3d} and
one of the referees for suggesting plotting the ratio $R$.
This work was
supported by the National Science Foundation under grant DMR 9713977.  M.P.
also is supported in part by a fellowship of Fondazione Angelo Della Riccia.
The numerical calculations were supported by computer time from the
National Partnership for Advanced Computational Infrastructure.

\end{document}